\documentclass[letterpaper,10pt]{article}

\usepackage{graphicx} % Required for inserting images
\usepackage{amsmath}
\usepackage{booktabs}%
\usepackage{placeins}
\usepackage{hyperref}
\usepackage{xcolor}
\usepackage{amsfonts} 
\usepackage{makecell}
\usepackage{biblatex} %Imports biblatex package
\title{Integrating LSTM Networks with Neural Lévy Processes for Financial Forecasting}

\author{
    \textbf{Mohammed Alruqimi}\thanks{Corresponding author: Mohammed Alruqimi, Affiliation, Department of Computer Science, University of
Verona,  Italy. Email: mohammed.alruqimi@univr.it}, 
    Luca Di Persio\thanks{Email: luca.dipersio@univr.it}
}
\author{
    \textbf{Mohammed Alruqimi}\thanks{Corresponding author: Email: mohammed.alruqimi@univr.it} \\
    Department of Computer Science, University of Verona, Italy \\
    \texttt{mohammed.alruqimi@univr.it} \\
    \\
    Luca Di Persio \\
    Department of Computer Science, University of Verona, Italy \\
    \texttt{luca.dipersio@univr.it}
}

\begin{document}
\date{}
\maketitle

\section{ Abstract}

This paper investigates an optimal integration of deep learning with financial models for robust asset price forecasting. Specifically, we developed a hybrid framework combining a Long Short-Term Memory (LSTM) network with the Merton-Lévy jump-diffusion model. To optimise this framework, we employed the Grey Wolf Optimizer (GWO) for the LSTM hyperparameter tuning, and we explored three calibration methods for the Merton-Lévy model parameters: Artificial Neural Networks (ANNs), the Marine Predators Algorithm (MPA), and the PyTorch-based TorchSDE library.
To evaluate the predictive performance of our hybrid model, we compared it against several benchmark models, including a standard LSTM and an LSTM combined with the Fractional Heston model.
This evaluation used three real-world financial datasets: Brent oil prices, the STOXX 600 index, and the IT40 index. Performance was assessed using standard metrics, including Mean Squared Error (MSE), Mean Absolute Error (MAE), Mean Squared Percentage Error (MSPE), and the coefficient of determination ($R^2$).
Our experimental results demonstrate that the hybrid model, combining a GWO-optimized LSTM network with the Lévy-Merton Jump-Diffusion model calibrated using an ANN, outperformed the base LSTM model and all other models developed in this study.

\section{Introduction}

One of the main research topics focused on financial econometrics is analysing asset dynamics and creating models that effectively capture asset returns’ dynamic patterns and characteristics.
Assets pricing models, fundamental in financial mathematics and risk management, have been extensively studied and recently enhanced through AI methodologies.
Numerous mathematical models have been proposed to capture price volatility or asset spikes that result from unexpected macroeconomic events or supply-demand imbalances \cite{Jimmy1998}.
Stochastic Volatility models, starting from the Black-Scholes model, are widely used in financial mathematics, aiding in pricing derivatives and risk management \cite{Dixit,Tsay2010}. 
The Black-Scholes model \cite{Black1973ThePO} has long been considered the foundational model for option pricing. 
However, it has faced criticism for its restrictive assumptions, such as constant volatility and the use of Geometric Brownian Motion (GBM), which do not accurately reflect the empirical characteristics of financial markets.
Numerous models have since been developed to relax the assumptions of the Black-Scholes model. Jump-diffusion models were introduced firstly by Merton \cite{MERTON1976125}, building on the option pricing work of Black and Scholes. Jump-diffusion is a time-dependent stochastic process, which includes both jumps and diffusion to replicate stylised facts observed in asset price dynamics, such as sudden price jumps and mean reversion. 
Heston introduced a stochastic volatility model \cite{heston93} with two Stochastic Differential Equations (SDEs), where the second SDE describes the volatility process assumed to follow an Ornstein-Uhlenbeck process.
Additionally, Lévy Processes \cite{levy} was developed to capture the skewness and kurtosis of return series, providing a more accurate representation of price dynamics.
Empirical analyses of models’ performance incorporating stochastic volatility and returns jumps can be found in \cite{BATES2000181,PAN20023,Ignatieva}.
 Many studies in energy and commodity markets emphasised the importance of incorporating stochastic volatility and jumps to capture market behaviour accurately. For instance,  \cite{LI2022106358,georgiev} applied jump-diffusion models to crude oil prices. \cite{LARSSON2011504} highlighted that jumps are essential for modelling crude oil price dynamics. \cite{BRIX2018560} advanced an Ornstein–Uhlenbeck-based model with stochastic volatility and Lévy processes. \cite{Ignatieva} presents a comprehensive exploration of applying stochastic volatility jump-diffusion models for the crude oil market dynamics. 
However, stochastic models are parametric and require appropriate parameter calibration for use. 
Financial model calibration implies adjusting model parameters to accurately reflect observed market data, such as prices or implied volatilities. This ensures that the model’s output aligns with real market behaviour and can be reliably used for forecasting and pricing derivatives. 
Recently, neural networks (NN) have been increasingly used for financial modelling calibration \cite{kim2023,Liu2019}, particularly for complex models with numerous parameters, to address accuracy, speed and robustness issues of the calibration.
The robustness of stochastic models lies in their transparency and the ability to validate and recalibrate them using financial principles and market data, ensuring consistent performance even during periods of market stress.
On the other hand, deep learning models excel in their adaptability and capacity to uncover complex, nonlinear relationships within large-scale data, often yielding superior predictive performance. However, deep learning approaches typically lack interpretability and are prone to overfitting, data biases, and limited generalisation when faced with changing market regimes or out-of-sample data \cite{Casolaro}. 
A popular recent solution involves leveraging these approaches to better capture patterns and modelling volatility (i.e. \cite{HU2020124907}).
In this paper, we explore various calibration techniques for calibrating the parameters of the Merton Jump Diffusion model and the Heston model for asset price prediction.
We also integrate three distinct approaches: Long Short-Term Memory (LSTM) networks, a Deep Merton Jump Diffusion model, and the Grey Wolf Optimizer (GWO) algorithm to enhance prediction accuracy and generalisation in asset price forecasting.

Despite the emergence of new models, LSTM remains a popular choice in financial price prediction. Developing more complex deep learning architectures can lead to overfitting, while advanced models like Transformers often require larger datasets \cite{Alruqimi2024}. Our approach centres on an ensemble forecast combining GWO-tuned LSTM predictions with a neural network-calibrated Merton Jump Diffusion model.

 The experiments showed the superiority of this proposed model against the benchmark models in terms of Mean Squared Error (MSE). 

\section{Related Works}

\subsection{Stocahstic-DL models for assets prices modelling}

Combining deep learning models with stochastic models for modelling asset price volatility and enhancing price prediction is a popular recent approach \cite{Qiguo2024,CHEN2021,HUANG2021116485,yan2020}. LSTM and GRU are among the most widely selected deep learning models in the literature \cite{Alruqimi2024}.
For instance, in a recent work, \cite{Qiguo2024} introduced a comprehensive framework for American option pricing that combines optimisation algorithms, numerical models, and neural networks to improve pricing accuracy and address data scarcity. The model employs transfer learning with numerical data augmentation and a jump diffusion-informed neural network to handle data scarcity, capture return distribution features, and use Bayesian optimisation to improve training efficiency.

\subsection{Financial models calibration}
In finance, asset model calibration involves estimating parameters of stochastic differential equations from market data. 
\cite{georgiev}  calibrated historical gold and crude Brent oil futures to the Merton jump-diffusion model using an improved log-return algorithm, which involves distinguishing between continuous movements and jumps through a threshold informed by Value at Risk (VaR).  \cite{Qiguo2024} used the Annealing algorithm to calibrate parameters of their framework that include a jump diffusion-informed neural network for American Option Pricing.
\cite{Liu2019} introduced an approach to calibrate financial asset price models using an Artificial Neural Network (ANN), mainly focusing on calibrating the Heston and Bates models.
\cite{kim2023} used ANNs to calibrate parameters for GARCH-type option pricing models: Duan’s GARCH and the classical tempered stable GARCH.

\section{Methodology}
This work integrates an LSTM network with financial models for asset price forecasting. We investigated two financial models: the Fractional Heston and the Lévy-Merton Jump-Diffusion models. Three methods were employed to calibrate the parameters of these financial models: Neural Networks, the Marine Predators Algorithm (MPA), and TorchSDE.
Figure \ref{fig:merton} illustrates the overall architecture framework. 

\begin{figure}[h]
    \centering
    \includegraphics[width=0.9\linewidth]{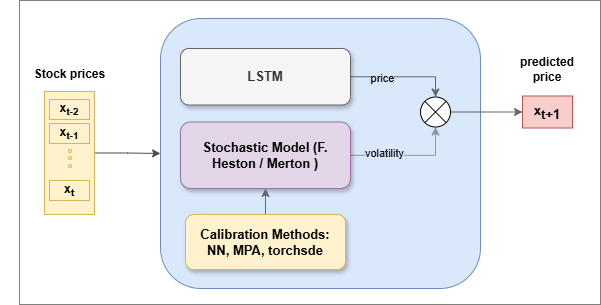}
    \caption{The proposed framework}
    \label{fig:enter-label}
\end{figure}

\subsection{ Fractional Heston Model}

The Heston model, introduced by Heston (1993) \cite{heston93}, characterises the dynamics of an underlying asset’s price and its variance through the following stochastic differential equations:

\[
    dS(t) = \mu S(t) \, dt + \sqrt{v(t)} S(t) \, dW_1(t)
\]

\[
    dv(t) = -\lambda (v(t) - \bar{v}) \, dt + \eta \sqrt{v(t)} \, dW_2(t)
\]

with initial conditions \( S(0) = S_0 \) and \( v(0) = v_0 \), where \( S(t) \) and \( v(t) \) represent the asset price and its variance at time \( t \), \( \mu \) is the drift rate of the asset, \( \lambda \) denotes the mean-reversion speed, \( \eta \) is the volatility of variance (volatility of volatility), and \( W_1(t) \) and \( W_2(t) \) are two correlated Brownian motions with correlation coefficient \( \rho \), i.e.,

The fractional Heston model extends the classical Heston model by introducing a fractional Brownian motion component to model the volatility. The dynamics of the asset price \( S_t \) and the variance \( v_t \) are described by:

\[
dS_t = \mu S_t dt + \sqrt{v_t} S_t dW_t^S,
\]
\[
dv_t = \kappa (\theta - v_t) dt + \xi v_t^\beta dW_t^v,
\]

Where \( W_t^S \) and \( W_t^v \) are two correlated Wiener processes with correlation coefficient \( \rho \), and \( \beta \) represents the degree of leverage effect, typically set to 0.5. The extended memory property is introduced through a fractional derivative in the volatility equation, represented by a fractional order \( H \) in the range \( (0.5, 1) \).

\subsection{Lévy Merton Jump-Diffusion Model}

The Merton Jump Diffusion model has been introduced by Merton’s paper \cite{MERTON1976125}.
The main idea of this paper was to extend the Black-Scholes model by introducing more realistic assumptions, addressing the fact that empirical studies of market returns do not adhere to a constant variance log-normal distribution. 
The Merton Jump Diffusion model is noteworthy for its ability to generate the volatility smile commonly observed in options markets. 
In the Merton Jump Diffusion model, the asset price $S_t$ follows the stochastic differential equation(SDE):

    \[
dS_t = S_t \left( (\mu - \lambda k) dt + \sigma dW_t + dQ_t \right)
\]

\begin{figure}[h]
     \centering
    \includegraphics[width=0.45\linewidth]{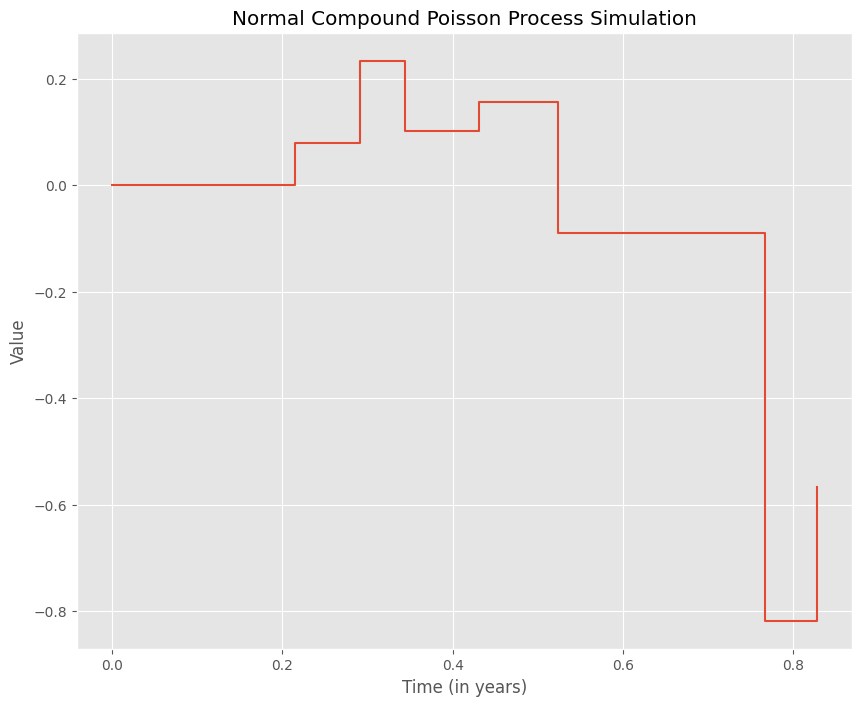}
     \includegraphics[width=0.48\linewidth]{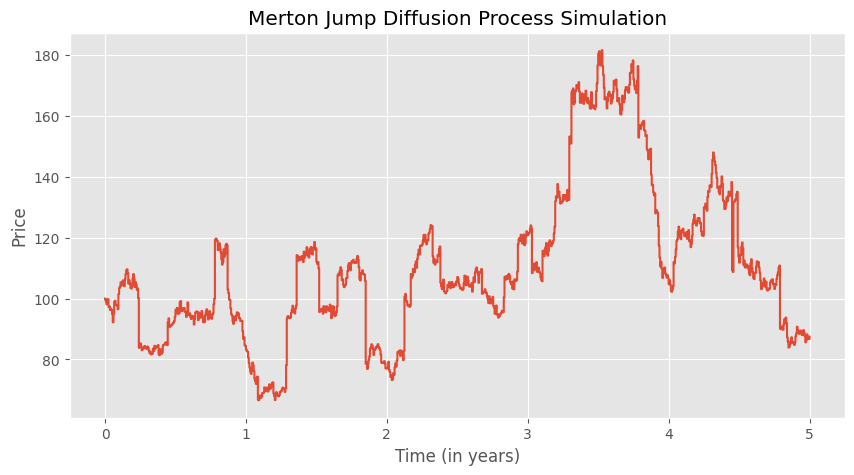}
     \caption{\textit{Left}: Normal compound Poisson process simulation.  $\lambda$ = 10, T = 1y, jumps $\sim$ N(0, 0.22). 10 jumps per year on average, the magnitude of the jumps follows a normal distribution N(0, 0.22). \\ \textit{Right}: Merton jump diffusion process simulation. $S_0$ = 100, $\mu$ = $5\%$, $\sigma$ = 20$\%$, $\lambda = 10, T = 5y$, jumps $\sim N(0, 0.12), \delta t = 5 / 1000$.}
     \label{fig:merton}
 \end{figure}

It contains two parts, Figure \ref{fig:merton}: 
\begin{itemize}
    \item Diffusion component: \( \sigma dW_t \), where \( W_t \) represents a standard Wiener process,
with \( \mu \) as the drift (expected return) and \( \sigma \) as the volatility.
This part captures the {\it continuous} price changes following a standard GBM.
\item Jump component: \( dQ_t \), modelled as a compound Poisson process to account for discontinuous jumps:

\[
dQ_t = \sum_{i=1}^{N_t} (Y_i - 1)
\]

where \( N_t \) is a Poisson process with intensity \( \lambda \), 
representing the number of jumps in the interval, and \( Y_i \) are i.i.d. random variables 
representing the jump sizes, which follow a log-normal distribution: \( \ln(Y_i) \sim \mathcal{N}(m, \delta^2) \). 

Accordingly, the term \( k = \mathbb{E}(Y_i - 1) = e^{m + \frac{1}{2}\delta^2} - 1 \) 
represents the expected relative jump size, and the parameter \( \lambda \) is the frequency of jumps. 
Let us underline that since the expectation of \( Q_t \) is:
\[
\mathbb{E}(Q_t) = \lambda k t
\]
then  \( Q_t - \lambda k t \) is a martingale, preserving the no-arbitrage condition of the model.
\end{itemize}

The Lévy–Merton jump-diffusion model is incorporated to enhance the forecast distribution by incorporating the jump dynamics of prices. 
During training, the jump-diffusion parameters (\( \mu, \sigma, \lambda, m, \delta \)) are optimised through a neural network gradient,  allowing the model to adjust to the complex volatility structure in real-time, which implies more accurate probabilistic forecasts and better identification of the risk of extreme price movements.
Integrating the Lévy–Merton model ensures a more realistic representation of price dynamics by handling continuous market fluctuations and discrete jumps.

\subsection{Calibration Methods}

We explored three methods for estimating the parameters of the Merton and Heston models: Neural Networks, the Marine Predators Algorithm (MPA), and TorchSDE.

\subsubsection{Neural Networks}
We utilise a feed-forward deep neural network with layers structured to encapsulate the nonlinear characteristics of the fractional Heston and Merton jump-diffusion models. The input layer accepts market prices, while the output layer provides estimates of the model parameters. For instances: \( \mu, \kappa, \theta, \xi, \rho, \) and \( H \) in the case of Heston model.
The neural network is designed as a deep, fully connected (dense) network chosen for its ability to model complex and nonlinear functions effectively. The architecture consists of two hidden layers, each employing the nonlinear activation function, the Rectified Linear Unit (ReLU).

\paragraph{Loss Function:}
The loss function is the mean squared error (MSE) between the market prices and those computed by the model using the NN-estimated parameters. Mathematically, it is expressed as:
\[
\mathcal{L}(\theta) = \frac{1}{N} \sum_{i=1}^N (P_{\text{market}, i} - P_{\text{model}, i}(\theta))^2,
\]
where \( P_{\text{model}, i}(\theta) \) denotes the model price computed with parameters \( \theta \).

\paragraph{Training and Optimization:}
The network is trained using the gradient descent method (Adam) to minimise the loss function. 
The input layer receives vectorised inputs consisting of observed market prices. This input vector is denoted by \( \mathbf{x} \in \mathbb{R}^d \), where \( d \) represents the dimensionality of the input data.
Several hidden layers process the input through weighted sums followed by nonlinear activations. Each layer \( l \) computes the following:
\[
\mathbf{z}^{(l)} = \sigma(\mathbf{W}^{(l)}\mathbf{a}^{(l-1)} + \mathbf{b}^{(l)}),
\]
where \( \mathbf{W}^{(l)} \) and \( \mathbf{b}^{(l)} \) are the weights and biases of the layer, \( \mathbf{a}^{(l-1)} \) is the activation from the previous layer (with \( \mathbf{a}^{(0)} = \mathbf{x} \)), \( \sigma \) is the activation function, and \( \mathbf{z}^{(l)} \) is the output of layer \( l \).
The output layer consists of a linear transformation that maps the final hidden layer’s activations to the parameter space of the fractional Heston or Merton jump-diffusion model:
\[
\boldsymbol{\theta} = \mathbf{W}^{(L)} \mathbf{a}^{(L-1)} + \mathbf{b}^{(L)},
\]
where \( L \) denotes the last layer of the network, and \( \boldsymbol{\theta} \) represents the estimated parameters. % \( (\mu, \kappa, \theta, \xi, \rho, H) \).

\paragraph{Gradient Computation:}
The gradient of the loss function concerning the network parameters is computed via backpropagation:
\[
\frac{\partial \mathcal{L}}{\partial \mathbf{W}^{(l)}} = \frac{\partial \mathcal{L}}{\partial \mathbf{z}^{(l)}} \cdot \frac{\partial \mathbf{z}^{(l)}}{\partial \mathbf{W}^{(l)}},
\]
\[
\frac{\partial \mathcal{L}}{\partial \mathbf{b}^{(l)}} = \frac{\partial \mathcal{L}}{\partial \mathbf{z}^{(l)}}.
\]
Here, the partial derivatives propagate errors from the output back to the input layer, adjusting the parameters to reduce prediction errors.

\paragraph{Parameter Update:}
After computing the gradients, parameters are updated using:
\[
\mathbf{W}^{(l)}_{\text{new}} = \mathbf{W}^{(l)} - \eta \frac{\partial
 \mathcal{L}}{\partial \mathbf{W}^{(l)}},
\]
\[
\mathbf{b}^{(l)}_{\text{new}} = \mathbf{b}^{(l)} - \eta \frac{\partial \mathcal{L}}{\partial \mathbf{b}^{(l)}},
\]
where \( \eta \) is the learning rate.
\subsubsection{TorchSDE Framework}

TorchSDE \cite{li2020scalable} is a PyTorch-based framework designed to streamline the integration of stochastic differential equations (SDEs) into machine learning workflows, enabling efficient gradient-based optimisation for stochastic systems. By leveraging PyTorch’s automatic differentiation and GPU acceleration, TorchSDE provides scalable tools for solving forward and backward SDEs. It suits high-dimensional problems common in modern machine learning applications, such as neural SDEs, stochastic control, and generative modelling. 

\subsubsection{Marine Predators Algorithm}
The Marine Predators Algorithm (MPA) \cite{FARAMARZI2020} is an optimisation technique inspired by the natural behaviours of marine predators during hunting, incorporating strategies from optimal foraging theory and the interaction rates between predators and prey. This metaheuristic is particularly effective in handling continuous optimisation problems. The algorithm mimics various phases of predation to navigate complex search spaces, balancing exploration and exploitation to enhance convergence toward optimal solutions.

 The mathematical formalisation involves dividing the search into three phases: high-velocity movement using a Lévy flight or Brownian motion during exploration, transitioning to a mixed phase with adaptive strategies, and a final exploitation phase for refining solutions.
 
During the initial optimisation phase in the Marine Predators Algorithm (MPA), exploration is emphasised when the predator moves faster than the prey. The mathematical model iterates until one-third of the maximum iterations is reached. It updates step sizes and positions using random vectors drawn from a normal distribution to mimic Brownian motion. This phase leverages entry-wise multiplication and random scaling to adjust prey positions, enhancing exploration capabilities for broad search coverage in the optimisation landscape.

\section{Experiments}
The experiments were conducted in two stages:
\begin{enumerate}
    \item Stage I: We trained an LSTM model for stock price forecasting in this step. Our objective was to use the LSTM model to generate accurate predictions for our dataset. We used the Grey Wolf Optimizer (GWO) to tune the model’s hyperparameters based on our previous work (\cite{Alruqimi2024-gwo}). The model’s performance was evaluated by calculating the Mean Squared Error (MSE) between predicted and actual prices.
    \item Stage II: This stage involved parameter calibration for two stochastic financial models: the Merton Jump-Diffusion model and the Fractional Heston model. The objective was to determine the optimal parameter values that best capture the volatility dynamics of each asset. Model calibration was performed using three distinct approaches: Neural Networks (NN), the Marine Predators Algorithm (MPA), and TorchSDE. The calibration accuracy of both model parameters was evaluated by integrating the generated volatility paths with Long Short-Term Memory (LSTM) forecasts. Subsequently, the Mean Squared Error (MSE) between the resulting ensemble forecast and the corresponding observed market prices was used as the performance metric.
\end{enumerate}

\subsection{Dataset}
To evaluate the proposed approach, we selected three real-world financial datasets covering different aspects of global economic markets:

\begin{enumerate}
    \item Brent Oil Price Dataset: This dataset tracks the daily prices of Brent crude oil, one of the key benchmarks for global oil pricing, from January 2010 to July 2024.
    \item STOXX 600 Dataset: Reflects the performance of the STOXX Europe 600 index, which includes 600 large, mid, and small-cap companies across 17 European countries. This dataset also includes historical observations from January 2010 to July 2024.
    \item IT40 Dataset: Contains data on Italy’s FTSE MIB index (IT40), the main benchmark index for the Italian equity market, covering its movement and trends during the study  (January 2010 to July 2024).
\end{enumerate}

These datasets encompass diverse market sectors, providing a comprehensive basis for evaluating the robustness and adaptability of the proposed approach in different financial contexts.

For data preparation, we applied two normalisation techniques using the scikit-learn package. The input feature $X$ was standardised using StandardScaler to ensure zero mean and unit variance, while the target variable $y$ was normalised to a range between 0 and 1 using MinMaxScaler.

\subsection{Calibration Performance}
Performance is quantified by comparing the calibrated parameters and the resulting model prices with market prices. Calibration error is defined as:
\[
\epsilon = \sqrt{\frac{1}{N} \sum_{i=1}^N (\tilde{C}_{\text{model}}(K_i, T_i; \boldsymbol{\theta}) - \tilde{C}_{\text{market}}(K_i, T_i))^2},
\]
Where \( \tilde{C}_{\text{model}} \) is the normalised model stock prices computed using the calibrated parameters.

\subsection{Computational Efficiency}
Computational efficiency is assessed by recording the training time and the number of iterations required to reach convergence. The improvement in computational time is compared to traditional methods, such as the Levenberg-Marquardt algorithm used in conventional calibration methods.

\subsection{Performance Metrics}
The predictive accuracy is assessed using several metrics:\\
- Mean Absolute Error (MAE):
  \[
  \text{MAE} = \frac{1}{N} \sum_{i=1}^N \left| C_{\text{predicted}}(K_i, T_i) - C_{\text{market}}(K_i, T_i) \right|
  \]
- Root Mean Squared Error (RMSE):\\
  \[
  \text{RMSE} = \sqrt{\frac{1}{N} \sum_{i=1}^N \left( C_{\text{predicted}}(K_i, T_i) - C_{\text{market}}(K_i, T_i) \right)^2}.
  \]
- R-squared (\(R^2\)):\\
  \[
  R^2 = 1 - \frac{\sum_{i=1}^N \left( C_{\text{predicted}}(K_i, T_i) - C_{\text{market}}(K_i, T_i) \right)^2}{\sum_{i=1}^N \left( C_{\text{market}}(K_i, T_i) - \overline{C}_{\text{market}} \right)^2},
  \]
  Where \( \overline{C}_{\text{market}} \) is the mean of the observed market prices in the test dataset.

\subsection{Model Prediction }
In this section, we compare results obtained from the developed models with different calibration methods. 
\subsubsection{LSTM-Lévy model}
This subsection presents results from the proposed hybrid LSTM-Merton-Lévy model for the three datasets.
\subsubsection*{Brent Oil Dataset}
\begin{figure}[h]
        \centering
        \includegraphics[width=0.45\linewidth]{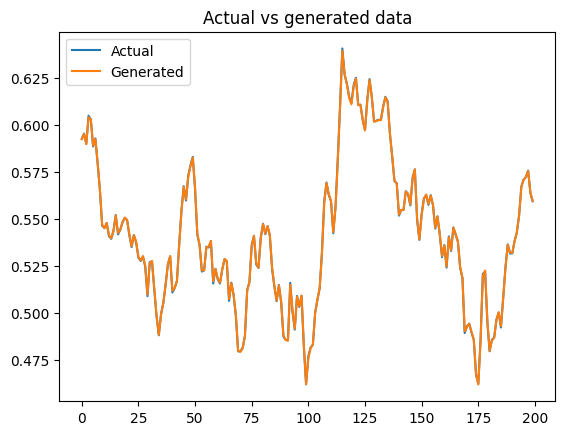}
         \includegraphics[width=0.45\linewidth]{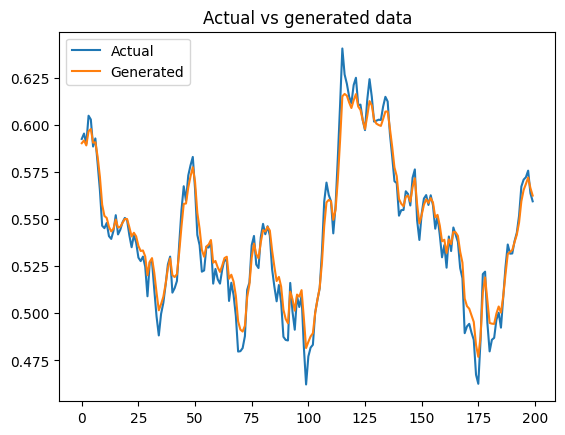}
         \includegraphics[width=0.45\linewidth]{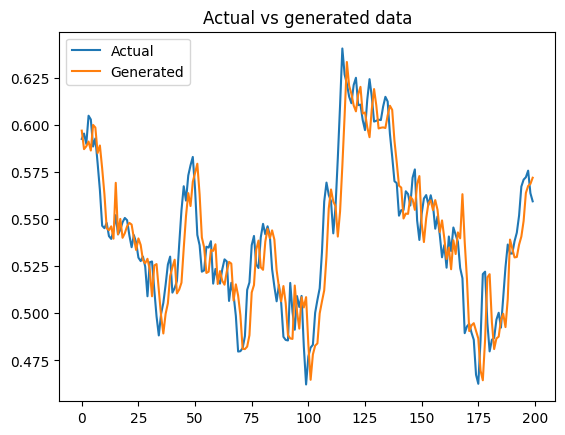}
         \includegraphics[width=0.45\linewidth]{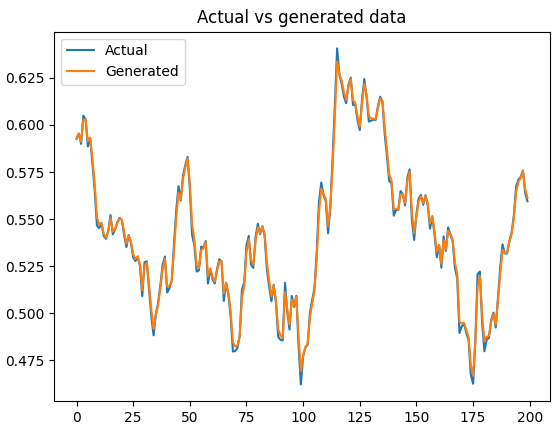}
        \caption{LSTM-Lévy model performance on the Brent Dataset.  The figures, arranged from top-left, present the results obtained using the three calibration methods: NN-based calibration, MPA-based calibration, and TorchSDE-based calibration of the Merton model, along with the performance of a standalone LSTM model.}
        \label{fig:merton_brent}
    \end{figure}

The evaluation metric for the Crude Brent oil dataset resulting from the application of the three calibration methods is presented in Table \ref{tab:eva_brent}.
While Table \ref{tab:sde_params_brent} shows the run time and values assigned to the Merton-Lévy parameters for the three calibration methods. 
The parameters are:
($\lambda$ ): intensity of Poisson process (intensity of jump);
($\mu$): drift;
m: mean of jumps; ($\sigma$): annual standard deviation, for Weiner process;
($\delta$): standard deviation of jumps.
Figure \ref{fig:merton_brent} plots the predicted Brent prices against the actual prices for 200 days. 

\begin{table}[h]
         \caption{Evaluation of the predictive performance of the LSTM-Lévy model for the Brent oil price dataset using three calibration methods for the Lévy parameters.}
         \label{tab:eva_brent}
         \setlength{\tabcolsep}{2pt} % Reduce column padding
         \begin{tabular}{lcccccc}
         \toprule
               Model & \makecell{Calibration \\ Method} & MAE & MSE & RMSE & MSPE & $R^2$\\ \midrule
              
              LSTM-Lévy & NN  & 0.0004500 & 0.0000045 & 0.0021213  & 0.0000152 & 0.996976\\
              LSTM-Lévy & torchsde & 0.0148499 & 0.0003764 &   0.0194036 &  0.0013148 & 0.7470060 \\
              LSTM-Lévy & MPA & 0.0050499 & 0.0000584 & 0.0076485 & 0.0002133 &  0.9606901 \\
              LSTM & - & 0.0021000 & 0.0000210 & 0.0045825 & 0.0000746 &  0.9858887 \\        
      \bottomrule
      \end{tabular} 
    \end{table}
\FloatBarrier
\begin{table}[!h]
         \caption{Lévy parameters estimated by the three methods and run time of the prediction models for the Brent oil prices dataset.}
         \label{tab:sde_params_brent}
         \begin{tabular}{lcccccc}
         \toprule
               Calibration \\ Method & run time & $\mu $ & $\sigma $  & $\lambda $ & m & $\delta$ \\ \midrule
              
              NN  & 240 & 0.58 & 0.05 & 0.19  & 0.0004 & 0.45\\ 
              torchsde  & 3158 & 0.046 & 0.0053 & 0.020 & 0.507 & 0.001 \\
              MPA & 2273 & 0.016 & 0.22 & 4.95 &  0.001 & 0.009\\ % job id 82
      
      \bottomrule
      \end{tabular} 
    \end{table}
\FloatBarrier
\FloatBarrier
\FloatBarrier

\subsubsection*{The STOXX Europe 600 dataset (STOXX 600)}

\begin{figure}[h]
        \centering
        \includegraphics[width=0.45\linewidth]{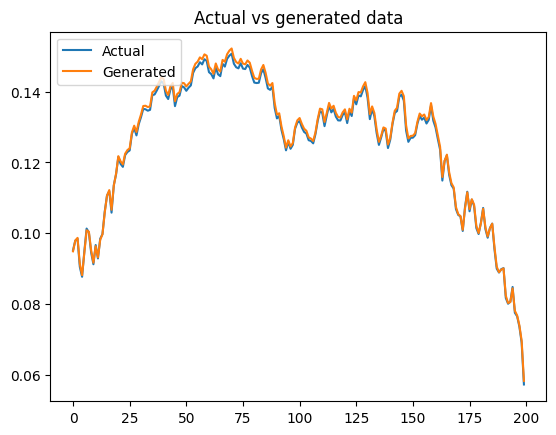}
         \includegraphics[width=0.45\linewidth]{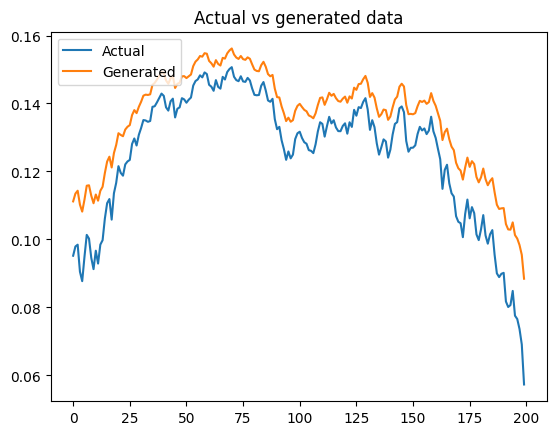}
         \includegraphics[width=0.45\linewidth]{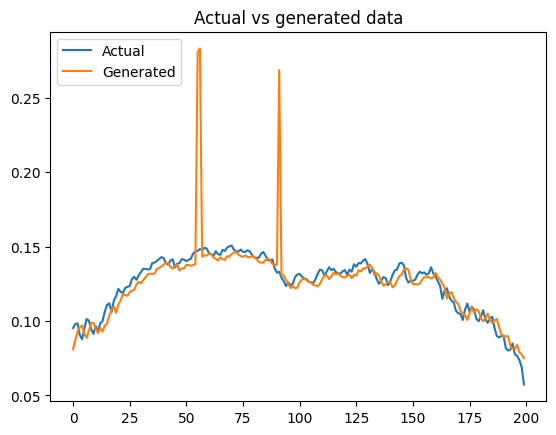}
         \includegraphics[width=0.45\linewidth]{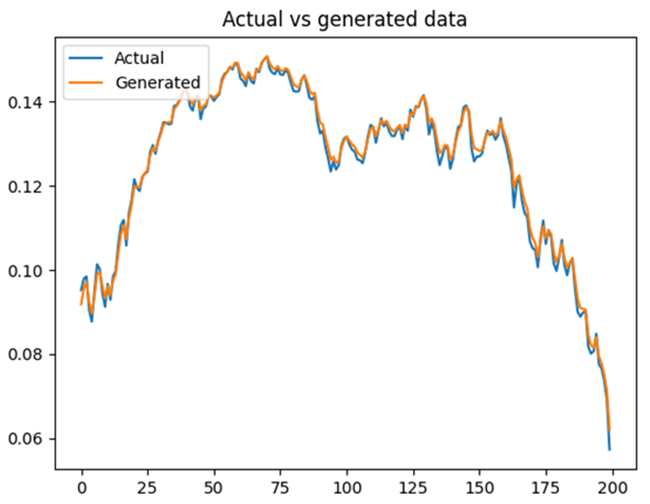}
        \caption{LSTM-Lévy model performance on the STOXX 600 Dataset.  The figures, arranged from top-left, present the results obtained using the three calibration methods: NN-based calibration, MPA-based calibration, and TorchSDE-based calibration of the Merton model, along with the performance of a standalone LSTM model.}
        \label{fig:merton_stoxx}
    \end{figure}

Table \ref{tab:eva_STOXX600} shows the results for the STOXX 600 dataset with the three calibration methods, while Table \ref{tab:sde_params_stoxx} shows the run time and Lévy parameters estimated by the three methods. Figure \ref{fig:merton_stoxx} plots the model predictions against the actual for this dataset.
\begin{table}[h]
         \caption{Evaluation of the predictive performance of the LSTM-Lévy model for STOXX 600 dataset using three calibration methods for the Lévy parameters.}
         \label{tab:eva_STOXX600}
         \setlength{\tabcolsep}{2pt} % Reduce column padding
         \begin{tabular}{lcccccc}
         \toprule
               Model & \makecell{Calibration \\ Method} & MAE & MSE & RMSE & MSPE & $R^2$\\ \midrule
              
              LSTM-Lévy & NN  & 0.0009000 & 0.0000090 & 0.0030000  & 0.0005877 &  0.9766209 \\
              LSTM-Lévy  & torchsde &  0.0066500 & 0.0003155 & 0.0177623 & 0.016960  & 0.1804342 \\ %job_id=102
              LSTM-Lévy  & MPA & 0.0072500 & 0.0000795 & 0.0089162 &  0.0050541 & 0.7934849  \\
             LSTM & - & 0.0012000 & 0.0000120 & 0.0034641 & 0.0008897 & 0.9688279 \\        
      \bottomrule
      \end{tabular} 
    \end{table}
\FloatBarrier

    \begin{table}[!h]
         \caption{Lévy parameters estimated by the three methods and run time of the prediction models for the STOXX 600 dataset.}
         \label{tab:sde_params_stoxx}
         \begin{tabular}{lcccccc}
         \toprule
               Calibration \\ Method & run time & $\mu $ & $\sigma $  & $\lambda $ & m & $\delta$ \\ \midrule
              
              NN  & 282 & 0.67 & 0.88 & 0.67 & 0.25 & 0.48 \\ 
              torchsde  & 1849 & 0.0552 & 0.0008 & 0.0200  & 0.000 & 0.0032 \\
              MPA & 3200 & 0.83 & 0.010 & 1.68  & 2.58  & 0.001\\ 
      
      \bottomrule
      \end{tabular} 
    \end{table}
\FloatBarrier
\FloatBarrier

\FloatBarrier

\subsubsection*{The Italy 40 dataset (IT 40)}
Table \ref{tab:eva_IT40} shows the evaluation metrics of the LSTM-Lèvy model for the ITALY 40 (IT 40) dataset with the three calibration methods. 

\begin{table}[h]
         \caption{Evaluation of the predictive performance of the LSTM-Lévy model for the Brent oil price dataset using three calibration methods for the Lévy parameters.}
         \label{tab:eva_IT40}
         \setlength{\tabcolsep}{2pt} % Reduce column padding
         \begin{tabular}{lcccccc}
         \toprule
              Model & \makecell{Calibration \\ Method} & MAE & MSE & RMSE & MSPE & $R^2$\\ \midrule
              
              LSTM-Lévy & NN  & 0.0006500 & 0.0000065 & 0.0025495 & 0.0000559 & 0.9981446 \\
              LSTM-Lévy & torchsde & 0.0200500  & 0.0008005 & 0.0282931 & 0.0059230  & 0.7715017 \\ 
              LSTM-Lévy & MPA & 0.0157000 & 0.0003530 &  0.018788 & 0.0028140  &  0.899238 \\
              LSTM & - & 0.0011500 & 0.0000115 & 0.00339116  & 0.0001028 & 0.996717 \\        
      \bottomrule
      \end{tabular} 
    \end{table}

    Table \ref{tab:sde_params_it40} shows the run time and Lévy parameters’ values estimated by the three methods for the IT 40 dataset.

    \begin{table}[h]
         \caption{ Lévy parameters estimated by the three methods and run time of the prediction models for IT 40 datase}
         \label{tab:sde_params_it40}
         \begin{tabular}{lcccccc}
         \toprule
               Calibration \\ Method & run time & $\mu $ & $\sigma $  & $\lambda $ & m & $\delta$ \\ \midrule
              
              NN  & 401 & 0.5583 & 0.0978 & 0.1980 & 0.00001 & 0.4051 \\ 
              torchsde & 1621 & 0.0423  & 0.0015 & 0.0200 & 0.000004 & 0.0006   \\
              MPA & 2771 & 0.020 &  0.092 & 4.98 & 0.00007  & 0.19\\ 
      
      \bottomrule
      \end{tabular} 
    \end{table}
    
\begin{figure}[h]
        \centering
        \includegraphics[width=0.45\linewidth]{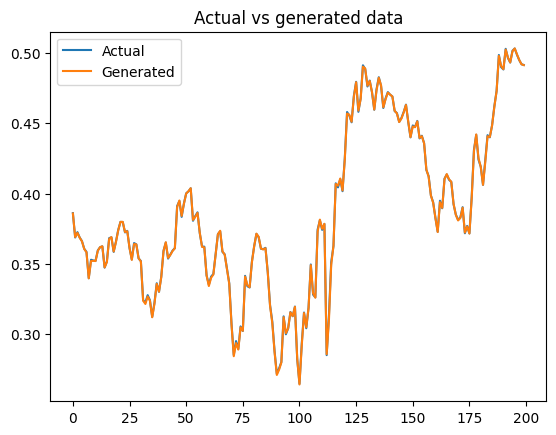}
         \includegraphics[width=0.45\linewidth]{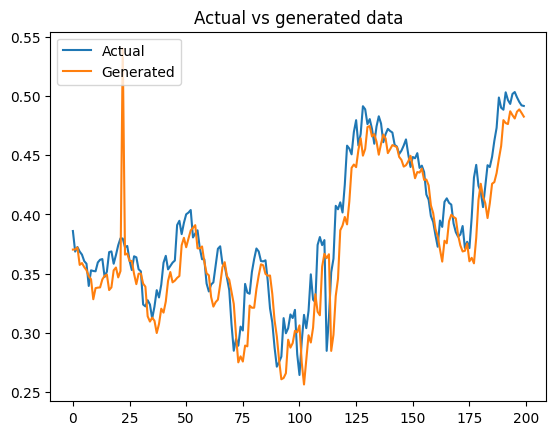}
         \includegraphics[width=0.45\linewidth]{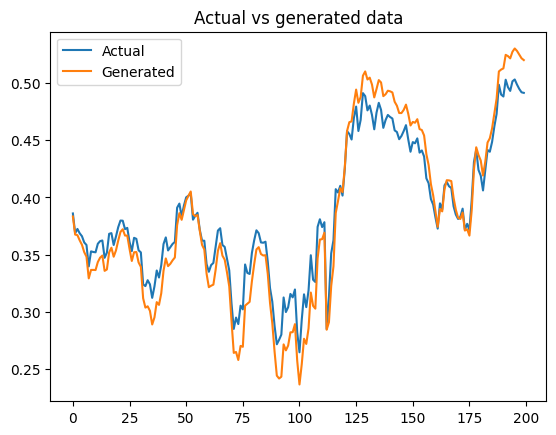}
         \includegraphics[width=0.45\linewidth]{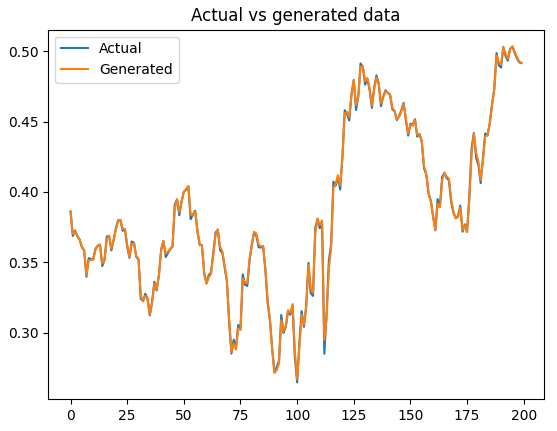}
        \caption{LSTM-Lévy model performance on the IT40 Dataset.  The figures, arranged from top-left, present the results obtained using the three calibration methods: NN-based calibration, MPA-based calibration, and TorchSDE-based calibration of the Merton model, along with the performance of a standalone LSTM model.}
        \label{fig:merton_it40}
    \end{figure}

Figure \ref{fig:merton_it40} shows the model predictions evaluations.
\FloatBarrier

\FloatBarrier

\subsubsection{LSTM-Fractional-Heston model}
In this section, we re-perform parts of the previous experiments using the Fractional Heston model instead of the Merton-Lévy model. In this regard, we implement our experiments with the NN-based calibration method.
Table \ref{tab:eva_hestont} shows the evaluation metrics for the LSTM  with the Fractional Heston model for the three datasets.
\FloatBarrier
\begin{table}[h]
         \caption{Evaluation of the predictive performance of the LSTM-Fractional-Heston model
         for the Brent oil price dataset. Fractional-Heston model parameters have been calibrated using the neural network method}
         \label{tab:eva_hestont}
         \begin{tabular}{lccccc}
         \toprule
               Dataset & MAE & MSE & RMSE & MSPE & $R^2$\\ \midrule
              
              Brent  & 0.0009000 &  0.0000090 &  0.003000 & 0.0000316 & 0.9939523 \\
              
              STOXX 600 & 0.0007000 & 0.0000070 & 0.0026457 & 0.0005189 &  0.9818162 \\
              
              IT 40  & 0.0007499 &  0.0000074 & 0.002738  & 0.0000651 & 0.997859 \\

      \bottomrule
      \end{tabular} 
    \end{table}
\FloatBarrier

\begin{figure}[h]
        \centering
        \includegraphics[width=0.32\linewidth]{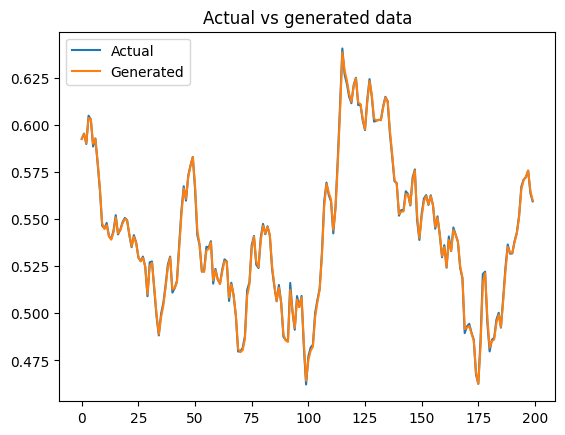}
         \includegraphics[width=0.32\linewidth]{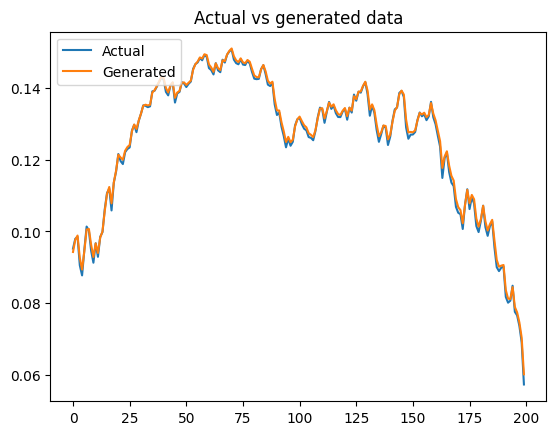}
         \includegraphics[width=0.32\linewidth]{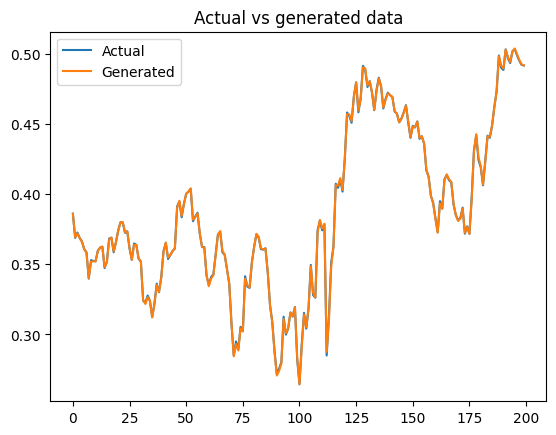}
        \caption{LSTM-Fractional-Heston model performance on the Brent Dataset. The figures, arranged from the top-left, present the results obtained using NN-based calibration for the Brent, STOXX 600, and ITALY 40 datasets.}

        \label{fig:heston_predictions}
    \end{figure}

Table \ref{tab:sde_heston_params} shows the Heston parameters estimated by NN, while Figure \ref{fig:heston_predictions} illustrates the actual versus generated values for each dataset.
\FloatBarrier
\begin{table}[h]
         \caption{The Fractional Heston parameters calibrated by NN}
         \label{tab:sde_heston_params}
         \begin{tabular}{lcccccc}
         \toprule
               Dataset & run time & $\mu $ & $\kappa $ & $\theta $ & $\varsigma$ & $\rho $ \\ \midrule
              
              Brent  & 302 & 0.0363 & 0.1474 & 0.2652  & 0.5233 & 0.1146 \\ 

              STOXX 600 & 358 & 0.0381 & 0.0779 & 0.2555 &  0.5131 & 0.1136   \\ 

              IT 40  & 394 & [0.0689 & 0.2038 &   0.2156 & 0.5245 & 0.1121\\ 
             
      \bottomrule
      \end{tabular} 
    \end{table}
\FloatBarrier
\FloatBarrier

\FloatBarrier

\section{ Discussion and Analysis}

The proposed framework demonstrated the efficacy of integrating Long Short-Term Memory (LSTM) models with neural network-calibrated financial models, particularly the Lévy–Merton Jump-Diffusion and Fractional Heston models. 

The results showed that LSTM combined with neural network-calibrated Lévy–Merton Jump-Diffusion consistently outperformed other calibration techniques, such as the Marine Predators Algorithm (MPA) and TorchSDE. Specifically, the neural network method achieved the lowest Mean Squared Error (MSE) and the highest R² values across the datasets.

The Fractional Heston model also performed well, particularly when neural networks were employed for calibration. But it slightly trailed behind the Lévy–Merton model in accuracy.

The neural network approach excelled in parameter calibration and computation efficiency compared to the other two methods. 

\subsection{Limitations and opportunities for improvements}
Further efforts can be investigated to enhance the approach of integrating stochastic jump models with deep learning for financial modelling. In particular, concerning $\alpha$-stable Lévy processes and tempered stable processes to provide richer dynamics and allow for jumps with varying magnitudes and frequencies,
hence yielding improved fit to empirical data. The class of $\alpha$-stable Lévy processes is particularly relevant 
when modelling phenomena that exhibit heavy tails, such as oil price returns. 
These processes are characterised by their infinite divisibility, 
essentially being defined by the Lévy-Khintchine representation, 
where the Lévy measure governing the jumps satisfies:

\[
\int_{\mathbb{R}} (1 \wedge |x|^2) \, \nu(dx) < \infty.
\]
Between them, we can consider those with stable distributions, 
which are parameterized by an index \( \alpha \in (0, 2] \). 
In particular, when \( \alpha < 2 \), the distribution of jumps exhibits power-law tails, 
which empirically aligns with the extreme price movements observed in oil markets.
In particular, for \( S_t \), the asset price, driven by an $\alpha$-stable Lévy process, we have the following Lévy-Ito decomposition:

\[
S_t = S_0 + \int_0^t b \, ds + \int_0^t \sigma \, dW_s + \int_0^t \int_{\mathbb{R}} x \, \tilde{N}(ds,dx),
\]

where \( \tilde{N}(ds, dx) \) is the compensated Poisson random measure associated with the Lévy process,
while the parameter \( \alpha \) governs the tail behavior, 
with smaller values of \( \alpha \) corresponding to heavier tails. 
For oil price data, which often exhibit significant kurtosis, we can find the optimal choice of \( \alpha \) 
by calibrating from historical returns. 
Moreover, the characteristic function of the stable process is given by:
\[
\mathbb{E}[e^{iuS_t}] = \exp\left( -t|u|^\alpha \left(1 - i\beta \, \text{sgn}(u) \tan\left(\frac{\pi \alpha}{2}\right)\right) + i \gamma u \right),
\]
where \( \beta \) controls skewness and \( \gamma \) is the location parameter,
and the absence of finite moments (for \( \alpha < 2 \)) implies that modelling higher moments, such as variance, must be done with care.
From a practical point of view, stable processes provide a more accurate representation of large price jumps while requiring advanced simulation techniques due to the lack of closed-form solutions for most stable distributions. 
Moreover, the tail index \( \alpha \) can be directly linked to the extremal index of the observed time series, enabling the modelling of extreme value theory within the Lévy framework.
While concerning the tempered stable process, which tempers the tails of a stable distribution, mitigating the infinite variance issue,
we should consider that they modify the Lévy measure of a stable process 
by introducing an exponential damping factor,
leading to a finite variance and improved numerical stability, 
indeed, the Lévy measure for a tempered stable process is given by:

\[
\nu(dx) = \frac{C}{|x|^{1+\alpha}} e^{-\lambda |x|} \, dx,
\]

where \( \lambda > 0 \) is the so-called tempering parameter,
while \( \alpha \in (0,2) \) controls the tail heaviness and the 
exponential tempering ensures that while the process retains heavy tails, the jumps are not as extreme as in the pure $\alpha$-stable case; accordingly, we have:
\[
S_t = S_0 + \int_0^t b \, ds + \int_0^t \sigma \, dW_s + \int_0^t \int_{\mathbb{R}} x \, \tilde{N}_{\lambda}(ds,dx),
\]
where \( \tilde{N}_{\lambda}(ds, dx) \) is the compensated Poisson random measure for the tempered stable process,
which is particularly useful in markets where extreme events occur but are tempered 
by regulatory or structural mechanisms, such as production adjustments in oil markets.
Also in this case, the parameters of the Lévy–Merton jump-diffusion model, 
specifically \( \sigma \) (volatility), \( \lambda \) (jump intensity), and \( k \) (jump size), 
are learned during the training of the neural network, the main challenge lies in separating the contributions 
of the diffusion component (the one driven by Brownian motion) 
and the jump one  (Poisson process or alternative jump structures). 
As we know in the Merton model, the diffusion component \( \sigma dW_t \) and the jump component \( dQ_t \) 
both contribute to the overall volatility of the price process, 
making it challenging to assign variations in the observed data uniquely to one or the other, especially when dealing with small datasets.
A potential solution involves regularising the estimation of the parameters by imposing constraints on the magnitude of jumps relative to the diffusion term. 
Namely,  we could introduce a penalty term in the loss function to enforce a balance between the two components:
\[
L(\theta) = L_{\text{original}} + \lambda_{\text{reg}} \left( \frac{\sigma}{k} \right),
\]
where \( L_{\text{original}} \) is the original loss (e.g., MSE) and \( \lambda_{\text{reg}} \) is a regularisation coefficient, hence ensuring that the volatility attributed to the jump component does not dominate the diffusion component without justification from the data. 
We can also state that regularisation can be imposed through 
a Bayesian hierarchical framework, where the parameters \( \lambda \) and \( k \) are modelled as random variables with prior distributions that encode prior knowledge about reasonable jump intensities and sizes, e.g., we could impose log-normal priors on these parameters, reflecting their positivity and typical scale in financial markets:
\[
\lambda \sim \text{LogNormal}(\mu_\lambda, \sigma_\lambda^2), \quad k \sim \text{LogNormal}(\mu_k, \sigma_k^2).
\]
and this Bayesian regularisation scheme naturally controls for overfitting by penalising extreme parameter values during training. 
Moreover, posterior inference can then be carried out using techniques such as Markov Chain Monte Carlo or Variational Inference, 
which would yield not only point estimates for \( \lambda \) and \( k \), but also credible intervals.

Other possible directions might be the subjects of future works to improve our findings further.
In particular, the calibration of the parameters \( \sigma \) , \( \lambda \) , and \( \delta \) characterising the Merton-lévy model is done jointly estimating them via neural networks. However, these parameters are not uniquely identifiable from asset price data alone. For instance, periods of high volatility could be attributed to either increased \( \sigma \) or elevated \( \lambda \), leading to confounding effects. This is evident in the likelihood function \( \mathcal{L}(\theta) \), which may exhibit multiple local maxima, rendering calibration unstable. In particular, since the observed price process \( S_t \) follows
\[
dS_t = S_t \left[ (\mu - \lambda k) dt + \sigma dW_t + dQ_t \right],
\]  
where \( dQ_t \), i.e. the {\it jump component}, is modelled as a compound Poisson process 
\[
dQ_t = \sum_{i=1}^{N_t} (Y_i - 1)
\]
\( N_t \) being a Poisson process with intensity \( \lambda \), 
representing the number of jumps in the interval. At the same time, \( Y_i \) are i.i.d. random variables identifying the jump sizes and following a log-normal distribution: \( \ln(Y_i) \sim \mathcal{N}(m, \delta^2) \), implying that  \( k = \mathbb{E}(Y_i - 1) = e^{m + \frac{1}{2}\delta^2} - 1 \) represents the expected relative jump size. The parameter \( \lambda \) is the frequency of jumps, then the quadratic variation \( [S]_t \) includes contributions from both terms:  
\[
[S]_t = \int_0^t \sigma^2 S_s^2 ds + \sum_{s \leq t} (\Delta S_s)^2,
\]  
making it challenging to disentangle \( \sigma \) from \( \lambda \) and \( \delta \) without additional constraints.
We aim to solve the latter issue  by introducing the Bayesian regularisation framework with informative priors to penalise implausible parameter combinations, e.g.:
\newline
(1) Assign conjugate priors \( \sigma \sim \text{Gamma}(a_\sigma, b_\sigma) \) and \( \lambda \sim \text{Gamma}(a_\lambda, b_\lambda) \), reflecting typical scales observed in financial markets;
\newline(2) Modify the loss function to include a Kullback-Leibler (KL) divergence term:  
\[
\mathcal{L}_{\text{new}}(\theta) = \mathcal{L}_{\text{MSE}}(\theta) + \gamma \left( \text{KL}(q(\theta) || p(\theta)) \right),
\]  
where \( p(\theta) \) is the prior and \( q(\theta) \) the posterior approximation. This penalises deviations from economically reasonable parameter values.  
We shall also employ regime-switching extensions to the above SDE, where parameters \( \sigma, \lambda \) adapt to market states, e.g., calm versus crisis periods. Accordingly, the SDE becomes:  
\[
dS_t = S_t \left[ (\mu - \lambda_{Z_t} k_{Z_t}) dt + \sigma_{Z_t} dW_t + dQ_t^{Z_t} \right],
\]  
with \( Z_t \in \{1, 2\} \) denoting latent states inferred via Hidden Markov Models (HMMs) or recurrent networks.
\newline
A second point we aim to consider for the extension of the present paper is linked to the fact that deemed datasets span 14 years (2010–2024), during which structural breaks, e.g., geopolitical crises and monetary policy shifts, potentially induced non-stationarities in asset price dynamics. The approaches we developed for the present paper preprocess data using StandardScaler and MinMaxScaler, but these linear transformations fail to address regime-dependent volatility or trending means. LSTMs, while capable of learning temporal dependencies, implicitly assume local stationarity within their memory horizon. For instance, a persistent upward trend in oil prices, e.g., post-2020 recovery, could bias forecasts if not explicitly modelled.
To overcome the latter issue, we are aiming to 
integrate fractional differencing into the LSTM architecture to handle non-stationarity. 
In particular, we will first define the differenced input \( \nabla^d X_t = X_t - \sum_{k=1}^\infty \frac{(-d)(-d+1)\cdots(-d+k-1)}{k!} X_{t-k} \), where \( d \in (0,1) \) is learned endogenously, hence extending  the LSTM cell to:  
\[
\begin{aligned}
f_t &= \sigma(W_f \cdot [\nabla^d X_t, h_{t-1}] + b_f), \\
i_t &= \sigma(W_i \cdot [\nabla^d X_t, h_{t-1}] + b_i), \\
\tilde{C}_t &= \tanh(W_C \cdot [\nabla^d X_t, h_{t-1}] + b_C), \\
C_t &= f_t \odot C_{t-1} + i_t \odot \tilde{C}_t, \\
o_t &= \sigma(W_o \cdot [\nabla^d X_t, h_{t-1}] + b_o), \\
h_t &= o_t \odot \tanh(C_t).
\end{aligned}
\]  
It is worth mentioning that the parameter \( d \) can be optimised via gradient descent alongside LSTM weights. Alternatively, we can also adopt a multiscale wavelet decomposition preprocessing step. Then, decomposing \( S_t \) into stationary subseries \( \{S_t^{(j)}\}_{j=1}^J \) using discrete wavelets:  
\[
S_t = \sum_{j=1}^J S_t^{(j)} + \epsilon_t,
\]  
and train separate LSTMs on each \( S_t^{(j)} \), we aim at isolating transient shocks, i.e. high-frequency components, from long-term trends, i.e. low-frequency components, enhancing model robustness.

\section{Conclusion}
This study introduced a hybrid deep learning and financial modeling framework for asset price forecasting, integrating an LSTM network with the Lévy-Merton Jump-Diffusion model. By leveraging the Grey Wolf Optimizer (GWO) for LSTM hyperparameter tuning and employing neural networks for financial model calibration, our approach achieved superior predictive performance. Experimental validation on Brent oil prices, the STOXX 600 index, and the IT40 index demonstrated that our hybrid model outperformed benchmark methods, including standard LSTM and LSTM-Fractional Heston models, across multiple error metrics.
We also discussed our findings, limitations, and opportunities for improvements in detail. 
The separation of diffusion and jump components remains challenging, particularly when handling small datasets, which can impact parameter estimation accuracy. Addressing these challenges through improved calibration techniques and more sophisticated stochastic modeling approaches could further refine the predictive capabilities of the proposed framework.
Additionally, while neural networks improve calibration efficiency, alternative approaches, such as Bayesian regularization or tempered stable processes, could further enhance robustness. Future research could explore the integration of $\alpha$ - stable Lévy processes to better capture heavy-tailed distributions in financial markets, as well as implement advanced regularization techniques to improve parameter interpretability and prevent overfitting. These enhancements would strengthen the model’s adaptability to diverse financial instruments and market conditions.
\printbibliography
\end{document}